\documentclass[journal,onecolumn]{IEEEtran}
\usepackage{amsfonts}
\usepackage{graphicx}      
\usepackage[subfigure]{graphfig}
\usepackage{subfigure}
\usepackage{booktabs}
\usepackage{amsmath}
\usepackage{amssymb}
\usepackage{multirow}
\usepackage{epstopdf}
\usepackage{float}
\usepackage[numbers,sort&compress]{natbib}

\begin{document}

\title{An improved mixture of probabilistic PCA for nonlinear data-driven process monitoring}

\author{~Jingxin Zhang,  Hao~Chen, Songhang Chen, and Xia Hong 
\thanks{Jingxin Zhang, Hao Chen and Songhang Chen are with the Department
of Quanzhou Institute of Equipment Manufacturing, Haixi Institutes, Chinese Academy of Sciences, Jinjiang 362200,
China. (e-mail: zjx18@mails.tsinghua.edu.cn).}
\thanks{Xia Hong is with the Department of Computer Science, School of Mathematical, Physical and Computational Sciences, University of Reading, RG6 6AY, U.K.}
\thanks{This paper has been published in IEEE Transaction on Cybernetics}}

\maketitle

\begin{abstract}
An improved mixture of probabilistic principal component analysis (PPCA)  has been introduced for nonlinear data-driven process monitoring in this paper. To realize this purpose, the technique of a mixture of probabilistic principal component analysers is utilized to establish the model of the underlying nonlinear process with local PPCA models, where a novel composite monitoring statistic is proposed based on the integration of two monitoring statistics in modified PPCA-based fault detection approach. Besides, the weighted mean  of the monitoring statistics aforementioned is utilised as a metrics to detect potential abnormalities. The virtues of the proposed algorithm have been discussed  in comparison with several unsupervised algorithms. Finally, Tennessee Eastman process and an autosuspension model are employed to demonstrate the effectiveness of the proposed scheme further.

\end{abstract}


\IEEEpeerreviewmaketitle

\section{Introduction}
For the sake of system reliability and operational safety, large-scale industrial systems and applications  increasingly demand improved process monitoring technologies, which have been extensively researched in recent decades~\cite{kwong2015fault,alizadeh2016a,wang2017adaptively,han2017robust}.
However,  often quantitative models are difficult to  establish  due to lack of prior knowledge. Alternatively sensing measurements which can replicate the desired process behavior are generally available and are utilized to design data-driven models~\cite{mu2017data-driven,li2015data-driven}. Thus, data-driven process monitoring techniques are becoming more prevalent and have been recognized as powerful tools for fault diagnosis purpose by comparison with knowledge-based approaches and approaches based on analytical model~\cite{hou2017overview,wang2016event,li2017real-time}.

Data-driven process monitoring methods have been extensively researched~\cite{yin2017fault,yin2014a}.  S.~X.~Ding has presented a comprehensive introduction on data-driven design of fault diagnosis and summarized several multivariate analysis techniques to fault diagnosis~\cite{data2014sx}, i.e., principal component analysis (PCA), dynamic PCA, partial least squares (PLS), canonical variate analysis (CCA), etc. Besides, benchmark applications such as three-tank system, continuous stirred tank heater, Tennessee Eastman process, are utilized to demonstrate the effectiveness of these approaches. Yin et al also summarized a variety of data-driven techniques for multivariate statistical process monitoring (MSPM), and analysed their computational complexities as well as data assumption~\cite{yin2012comparison}, including independent component analysis (ICA), fisher discriminant analysis (FDA), subspace aided approach, etc.  However, these data-driven techniques are mainly applied to linear systems. Aimed at nonlinear applications, various variants of these basic approaches are developed, e.g.,  standard techniques under Gaussian model or probabilistic model~\cite{queiroz2016fault,kristan2014online,pica,lu2015online}.

Amongst many data-driven approaches, PCA serves as a classical technique for feature extraction due to its simplicity and effectiveness~\cite{wang2016principal,yin2015online}. Therefore, PCA has been widely employed for MSPM in recent years~\cite{disPCA,pca1,pcamm}. Several extensions of traditional PCA have been proposed to settle different issues, e.g., parameter variation~\cite{rpca},  practical batch process~\cite{batchPCA}, large-scale process~\cite{zhu2017distributed}, detecting slowly developing drifts~\cite{mwpca}.

However, traditional PCA scheme still has various limitations, e.g.  the basic assumption of multivariate Gaussian distributed data. Besides, the efficiency and detectability of PCA-based technique would be greatly discounted for nonlinear applications.  Numerous sophisticated variants of PCA have been intensively studied to tackle these problems. For instance, the locally weighted projection regression (LWPR) is a nonlinear regression method~\cite{yin2016improved}, where PCA-based process monitoring model can be computed under the locally weighted framework~\cite{yin2017data}.  In~\cite{cheng2005nonlinear,chen2002dynamic},  just-in-time learning (JITL) or neural network (NN) serves as process model to account for the nonlinear as well as dynamic behavior of the process, followed by  PCA which  analyzes the residuals from the difference between predicted outputs and process outputs.  In addition, PCA under Gaussian mixture model can be free from the assumption of Gaussian distribution~\cite{gmpca}.  In work of ~\cite{Chemical,alcala2010reconstruction,ge2009improved}, kernel PCA (KPCA) is mainly aimed at tackling nonlinearity problem. Furthermore, other forms of nonlinear PCA are also discussed in ~\cite{liu2011improved,iacoviello2016a}.
Nevertheless, the effectiveness of KPCA is largely dependent on the option of kernel functions and the corresponding critical parameters. In addition, traditional PCA algorithm performs badly when data values are incomplete. Aimed at this problem, probabilistic PCA (PPCA) was proposed, where expectation maximization (EM) algorithm could estimate the principal subspace iteratively~\cite{tipping1999probabilistic}. Note that the applications of PPCA should satisfy the basic assumption that process data follow multivariate Gaussian distribution.

In order to  deal with nonlinearities that are inherent in many underlying systems, it is of practical interest to integrate multiple PCA models to get complicated projection schemes~\cite{multiPCA}. Taking the virtues of PPCA aforementioned into consideration, a complicated model is readily implemented as a combination of such PPCA models via the technique of a mixture of probabilistic principal component analysers, namely, MPPCA approach~\cite{MPPCA,zhao2014efficient}.  MPPCA enables dealing with any probability density function and can figure out the global linearity of PCA. Generally speaking, MPPCA inherits the benefits of PPCA and can be applied to nonlinear systems.

Due to virtues aforementioned, MPPCA has been utilized for process monitoring~\cite{sharifi2017nonlinear,ge2010mixture}. In \cite{sharifi2017nonlinear}, MPPCA is utilized for sensor fault diagnosis purpose, but detailed theory of fault detection logic is unavailable. A mixture Bayesian regularization method of PPCA was proposed for multimode process monitoring~\cite{ge2010mixture}. However, it can not be applied to nonlinear and non-Gaussian systems. Therefore, it is valuable and necessary to present a specific description about improved MPPCA approach  for nonlinear process monitoring.

In this paper,  we proposed an improved nonlinear data-driven process monitoring algorithm based on MPPCA, referred to as I-MPPCA.  A new monitoring statistic is introduced based on the integration of two monitoring statistics in modified PPCA-based fault detection approach. Besides, the weighted mean of the monitoring statistics aforementioned is developed to detect potential abnormalities.  The major advantages of I-MPPCA are summarized as follows:

1) It can cope with the global linearity of PCA, which is appropriate and effective to monitor nonlinear process;

2) The weight of a new data point belonging to a certain local PPCA model can be interpreted by the probability of each local model of being chosen;


3) It can process with any probability density function and owns lower computational complexity as well as stronger parameter robustness than kernel approaches;

4) It can deliver an optimal monitoring performance even when some data values  are missing;

5) Compared with traditional MPPCA technique, only one or two global monitoring statistics are developed for process monitoring, which is more convenient for practical industrial applications.

The rest of this paper is organized below. Section~\ref{basic_theory} reviews concepts and mathematical formulations of the probabilistic PCA and MPPCA as preliminaries of our proposed approach. Section~\ref{proposed_approach} details the proposed I-MPPCA approach, in which we propose model selection for I-MPPCA, novel monitoring statistics as well as corresponding thresholds. Besides, the monitoring performance is discussed between the proposed approach and the existing unsupervised approaches. In Section~\ref{TE_CASE}, Tennessee Eastman (TE) process is employed to illustrate the rationality and virtues of the proposed approach in contrast with traditional MPPCA scheme. Then, an autosuspension model is adopted to demonstrate the superiorities of the proposed approach in comparison with other unsupervised schemes in Section~\ref{autosuspension_case}. Concluding remarks are given in Section~\ref{conclusion}.

\section{Preliminaries}\label{basic_theory}

\subsection{Latent variable models and PCA}\label{latent}
Consider using  the following model
\begin{equation}\label{Eq1}
  {\boldsymbol t} = {\boldsymbol y}({\boldsymbol x};{\boldsymbol w}) + {\boldsymbol \xi}
\end{equation}
 to describe the observation  process vector ${\boldsymbol t}\in \Re^{d}$, where  $ \boldsymbol{x}\in \Re^{q}$  is the vector of  latent   variables, ${\boldsymbol w}$ is the associated model parameter vector.
$ \boldsymbol{\varepsilon}$ is an independent noise  vector.   $ \boldsymbol{y}\left(   \boldsymbol{x};{\boldsymbol w}   \right) $ is the unknown function of the system. For example, it   can be interpreted by a linear model used in statistical factor analysis, given by
\begin{equation}\label{Eq2}
   \boldsymbol {t} = {\boldsymbol W} {\boldsymbol x}  +  \boldsymbol{\mu}+ {\boldsymbol \xi}
\end{equation}
By means of defining a prior distribution over ${\boldsymbol x}$ and  of ${\boldsymbol \xi}$, a related distribution is induced in the data space according to (\ref{Eq2}).  Then, maximum-likelihood approach is utilized to determine the model parameters given a set observational data.


We assume ${\boldsymbol x} \sim {\rm N}(\boldsymbol{0},{\boldsymbol I})$, ${\boldsymbol  \xi }  \sim {\rm N}(\boldsymbol{0},\boldsymbol{\Psi} )$.  $\boldsymbol{0}$ and $\boldsymbol{I}$ denote the vector of all zeros and identity matrix  with appropriate dimensions respectively. $\boldsymbol{\Psi}  \in \Re^{ d \times d  }$ is assumed to be a diagonal matrix, $\boldsymbol{\mu} \in \Re^{d}$ is the output mean  vector, ${\boldsymbol W} \in \Re ^{d \times q}$ is the matrix of loading factors. Based on (\ref{Eq2}), it can be shown that the observation vector obeys Gaussian distribution ${\boldsymbol t} \sim {\rm N}({\boldsymbol \mu} ,{\boldsymbol C})$ with
 ${\boldsymbol C}= {\boldsymbol  \Psi}  + {\boldsymbol W}{\boldsymbol W}^{\rm T} \in \Re^{d \times d}$.

 Consider that a data set of $N$ output data samples $\{{\boldsymbol t}_n  \}_{n=1}^N$ is available. Over the data set,  (\ref{Eq2}) can be represented in matrix form as
\begin{equation}
\boldsymbol {T}={\boldsymbol W} {\boldsymbol X}+\boldsymbol{u}\boldsymbol{1}^{\rm T}+\boldsymbol {\Xi}
\end{equation}
where $\boldsymbol {T}=[\boldsymbol {t}_1,..., \boldsymbol {t}_N] \in \Re^{d \times N}$, $\boldsymbol {X}=[\boldsymbol {x}_1,... \boldsymbol {x}_N] \in \Re^{q \times N}$, $\boldsymbol {\Xi}=[\boldsymbol {\xi}_1,... \boldsymbol {\xi}_N ] \in \Re^{d \times N}$. $\boldsymbol{1}$ denotes the vector of all ones with appropriate dimension.
Let $\boldsymbol{u}=\frac{1}{N}\sum_{n=1}^{N}\boldsymbol {t}_n$, and the sample covariance matrix be denoted by ${\boldsymbol S}= \frac{1}{N} (\boldsymbol {T}-\boldsymbol{u}\boldsymbol{1}^{\rm T})(\boldsymbol {T}-\boldsymbol{u}\boldsymbol{1}^{\rm T})^{\rm T}$. There exist certain links between factor analysis and PCA, which have been demonstrated in~\cite{estimation}.  PCA problem can be settled by factor analysis.

Denote the eigenvalue decomposition of ${\boldsymbol S}=   \boldsymbol  {\tilde{W}}  {\boldsymbol \Lambda}  \boldsymbol {\tilde{ W}}^{\rm T}$, where ${\boldsymbol \Lambda}={\rm diag}\{\lambda_1,...,\lambda_q, \lambda_{q+1},...,\lambda_d \}$,
 $ \lambda_1>\lambda_2>...>\lambda_q>0$  are   nonzero eigenvalues ${\boldsymbol S}$, the $d-q$ smallest eigenvalues are minor and negligible.  Specifically we find $q$-dimensional vectors
\begin{equation}
{\boldsymbol x}_n= {\boldsymbol W}^{\rm T}(\boldsymbol {t}_n-\boldsymbol{\mu})
\end{equation}
to represent ${\boldsymbol t}_n$, $n=1,...N$, where ${\boldsymbol W}$ is the first $q$ columns of ${\boldsymbol {\tilde W}}$ and $q$ is determined by accumulating contribution rate. It can be shown that projection onto  most dominant  eigenvectors  leads to finding lower dimensional latent variables whilst retaining  maximal variance  of the original variables $\boldsymbol t_n$.

\subsection{The probabilistic PCA}
Providing that noise $ {\boldsymbol \varepsilon} \sim {\rm N}({\boldsymbol 0},{\sigma ^2}\boldsymbol {I})$, a probability distribution over $\boldsymbol t$-space is revealed for a specific $\boldsymbol x$ by the following formula
\begin{equation}\label{eq3}
  p({\boldsymbol t}|{\boldsymbol x}) = {(2\pi {\sigma ^2})^{ - d/2}}\exp \left\{ { - \frac{1}{{2{\sigma ^2}}}{{\left\| {\boldsymbol t - {\boldsymbol W} {\boldsymbol x} - \boldsymbol \mu } \right\|}^2}} \right\}
\end{equation}

A Gaussian prior probability over $\boldsymbol x$ can be defined by
\begin{equation}\label{eq4}
  p(\boldsymbol x) = {(2\pi)^{ - q/2}}\exp \left\{ { - \frac{1}{2}{{\boldsymbol x}^{\rm T}} {\boldsymbol x}} \right\}
\end{equation}

Then, the marginal distribution of $\boldsymbol t$ can be acquired in the form of
\begin{equation}\label{eq5}
\sloppy
\begin{array}{l}
p(\boldsymbol t) = \int { p({\boldsymbol t}|{\boldsymbol x})p(\boldsymbol x)d{\boldsymbol x}}\\
\quad \quad  = {(2\pi )^{ - d/2}}{\left| { \boldsymbol C} \right|^{ - 1/2}}\exp \left\{ { - \frac{1}{2}{{({\boldsymbol t} - \boldsymbol \mu )}^{\rm T}}{{\boldsymbol C}^{ - 1}}(\boldsymbol t - \boldsymbol \mu )} \right\}
\end{array}
\end{equation}
where $\left|  \cdot  \right|$ denotes matrix determinant. The model covariance is given by
\begin{equation}\label{eq6}
  \boldsymbol C = {\sigma ^2}{\boldsymbol I} + \boldsymbol W{{\boldsymbol W}^{\rm T}}
\end{equation}

In accordance with Bayesian theory, given the observation vector $\boldsymbol t$, the corresponding posterior distribution of $\boldsymbol x$ may be calculated:
\begin{align}\label{eq7}
& p(\boldsymbol x|\boldsymbol t) =  \rm{exp} \left\{ -\frac{1}{2} \left\{\boldsymbol x-{\boldsymbol M^{ - 1}}{\boldsymbol W^{\rm T}}(\boldsymbol t -\boldsymbol \mu ) \right\} ^{\rm T} {({\sigma ^{ - 2}}{\boldsymbol M})} \right. \nonumber\\
&\;\;\;\;\;\;\;\;\;\;\;\;\left. { \left\{\boldsymbol x-{\boldsymbol M^{ - 1}}{\boldsymbol W^{\rm T}}(\boldsymbol t -\boldsymbol \mu ) \right\}} \right\} \times  (2 \pi)^{-q/2} \left|{\sigma ^{ -2}}{\boldsymbol M}\right|^{1/2}
\end{align}
where the posterior covariance satisfies
\begin{equation}\label{eq8}
  {\sigma ^2}{\boldsymbol M^{ - 1}} = {\sigma ^2}{({\sigma ^2}{\boldsymbol I} + {{\boldsymbol W}^{\rm T}}{\boldsymbol W})^{ - 1}}
\end{equation}
with $\boldsymbol{M}  \in \Re^{ q\times q }$, $\boldsymbol{C}  \in \Re^{ d\times d  }$ .

The log-likelihood of the observation vector in this model is
\begin{equation}\label{eq9}
\begin{array}{l}
L = \sum\limits_{n = 1}^N {\ln \left\{ {p\left( {{{\boldsymbol t_n}}} \right)} \right\}} \\
\;\;{\kern 1pt}  =  - \frac{N}{2}\left\{ {d\ln \left( {2\pi } \right) + \ln \left|{\boldsymbol C } \right| + tr\left( {{{\boldsymbol C}^{ - 1}}{\boldsymbol S}} \right)} \right\}
\end{array}
\end{equation}

It can be shown~\cite{MPPCA} that  the log-likelihood (\ref{eq9})  is maximized when the columns of ${\boldsymbol W}$ span the principal subspace of the data. Analytical solutions can also be obtained via  the eigen-decomposition of ${\boldsymbol S}$, together with the estimation of noise variance $\sigma ^2$ (based on the smallest eigenvalues of ${\boldsymbol S}$). Alternatively iterative method of the EM algorithm can be used to generate the following complete-data log-likelihood as
\begin{align}
&L_c=\sum_{n=1}^{N} \ln\left\{p({\boldsymbol t}_n, {\boldsymbol x}_n ) \right\} \nonumber\\
&\;\;\;\; = \sum_{n=1}^{N} \ln\left\{{(2\pi {\sigma ^2})^{ - d/2}}\exp \left\{ { - \frac{1}{{2{\sigma ^2}}}{{\left\| {\boldsymbol t}_n - {\boldsymbol W} {\boldsymbol x}_n - {\boldsymbol \mu}   \right\|}^2}} \right\} \right. \nonumber\\
&\;\;\;\; \;\;\;\; \left. {(2\pi )^{ - q/2}}\exp \left\{ { - \frac{1}{2} {\boldsymbol x}_n^{\rm T}  {\boldsymbol x}_n}  \right\}\right\}
\end{align}
For convenience, we point out that the following qualities are used in EM algorithm~\cite{MPPCA}
 \begin{equation}\label{eqx}
   \left\langle {{\boldsymbol x_{n}}} \right\rangle  = {\boldsymbol M }^{ - 1}{\boldsymbol W }^{\rm T} \left( {{\boldsymbol t_n} - {\boldsymbol \mu }} \right)
 \end{equation}
\begin{equation}\label{eqxx}
  \left\langle {{\boldsymbol x_{n}}{\boldsymbol x_{n}^{\rm T}}} \right\rangle  = \sigma  ^2 {\boldsymbol M }^{ - 1} + \left\langle {{\boldsymbol x_{n}}} \right\rangle {\left\langle {{\boldsymbol x_{n}}} \right\rangle ^{\rm T}}
\end{equation}
denoting the expected posterior mean and covariance vectors  based on the latent model (\ref{Eq2}).

\subsection{The mixture of probabilistic PCA}
In order to be able to model more complex data, the mixture of probabilistic PCA (MPPCA) has been introduced~\cite{MPPCA},  which takes advantage of an integration of local PCA models via defining a mixture of probabilistic densities on the predicted output from $K$ local PCA models~\cite{MPPCA}. Our proposed I-MPPCA approach for process monitoring is based on MPPCA since it provides a more powerful base to handle nonlinearities and  missing data.

Instead of representing the system (\ref{Eq1}) by a single model of (\ref{Eq2}), we will consider, in accordance to probability rules,  a mixture of $K$ local PCA models,   as
\begin{align}\label{pt}
&p(\boldsymbol{t})= \sum_{i=1}^{K}p(i)p(\boldsymbol{t}|i)\nonumber\\
&\;\;\;\;\;\;=\sum_{i=1}^{K}\pi_ip(\boldsymbol{t}|i))
\end{align}
where the constraints on the mixing coefficients are ${\pi _i} \ge 0$ and $\sum {{\pi _i} = 1} $. $p(i)$ is interpreted as probability of choosing the $i$th local model.  Each of $p(\boldsymbol{t}|i)$ is  the local PCA model given by
\begin{equation}
   \boldsymbol {t} = {\boldsymbol W}_i {\boldsymbol x}   +  \boldsymbol{\mu}_i+ {\boldsymbol \xi}_i, \ \ i=1, ..., K
\end{equation}
which is similar to (\ref{Eq2}), and has individual projection matrice ${\boldsymbol W}_i$, mean vector $\boldsymbol{\mu}_i$, as well as  $ {\boldsymbol \xi}_i \sim {\rm N}({\boldsymbol 0},{\sigma_i ^2}\boldsymbol {I})$.

By comparing with equations (\ref{eqx}) and (\ref{eqxx}), the expected posterior mean and covariance vectors based on each of $K$ local PCA can be evaluated as
 \begin{equation}
   \left\langle {{\boldsymbol x^{(i)}_{n}}} \right\rangle   = {\boldsymbol M }_i^{ - 1}{\boldsymbol W }_i^{\rm T}\left( {{\boldsymbol t_n} - {\boldsymbol \mu_i }} \right) \in \Re ^{q}
 \end{equation}
\begin{equation}
  \left\langle   {{\boldsymbol x^{(i)}_{n}}}  {{\boldsymbol x^{(i)}_{n}}}^{\rm T}   \right\rangle   = \sigma_i  ^2 {\boldsymbol M }_i^{ - 1} + \left\langle {{\boldsymbol x^{(i)}_{n}}} \right\rangle  {\left\langle {{\boldsymbol x^{(i)}_{n}}} \right\rangle ^{\rm T}} \in \Re ^{q\times q}
\end{equation}
and are given here for convenience. Note that  ${{\boldsymbol x^{(i)}_{n}}}$ is $n$th sample for each $i$th model.

The solution procedure of MPPCA is presented in Appendix \ref{mppcasolution}, where two-stage EM scheme is adopted to improve convergence speed and reduce computational cost in Appendix \ref{solutionofmppca}.

\section{The proposed I-MPPCA approach for process monitoring}\label{proposed_approach}

\begin{table*}[!hbpt]
\caption{Overall procedure of I-MPPCA approach for process monitoring} \label{table1}
\begin{center}
\renewcommand{\arraystretch}{1.1}
\begin{tabular}{l}
\hline
\hline
\textbf{Phase 1: Off-line learning}\\
~~1. Collect $N$ normal sensing measurements from a  nonlinear process. \\
~~2. For I-MPPCA with $K$ local models\\
~~3. Estimate the model parameters.\\
~~~~~3a. Initialize the model parameters $\boldsymbol W_i$ and $\boldsymbol \mu_i$ ($i = 1:K$) via traditional PCA.\\
~~~~~3b. Calculate the number of principal components through accumulation contribution rate.\\
~~~~~3c. Discover the optimal model parameters ${\boldsymbol \theta}  = {\left\{ {{\boldsymbol W_i},{\boldsymbol \mu _i},\sigma _i^2,{\pi _i}} \right\}_{i = 1:K}}$ by implementation of the two-stage EM schedule.\\
~~~~~3d. Calculate $H(K)$ and go to step $2$ until $K={K_{\max }}$.\\
~~4. Discover the optimal number of local models $ {K^ \circ }$ among $K = \left\{ {1,2, \ldots ,{K_{\max} }} \right\}$ via the criterion given in (\ref{mn}).\\
~~5. Calculate the global monitoring statistics from equations (\ref{t22})-(\ref{tc2}) under normal operation condition.\\
~~6. Calculate the thresholds of the monitoring statistics from (\ref{jt2}) based on the calculated statistics aforementioned.\\ 
\textbf{Phase 2: On-line monitoring}\\
~~1. Get new sample and preprocess input data according to ${\boldsymbol \mu}_i$ ($i = 1:{K^ \circ }$) aforementioned.\\
~~2. Calculate local monitoring statistics in each local model according to equations (\ref{t2})-(\ref{tc}).\\
~~3. Calculate the global monitoring statistics from equations (\ref{t22})-(\ref{tc2}).\\
~~4. Detect faults according to the fault detection logic according to (\ref{fa1}) or (\ref{fa2}).\\
\hline
\hline
\end{tabular}
\end{center}
\end{table*}

The major procedure of MPPCA-based fault detection scheme can be summarized thereinafter. First, the model framework including the basic information of local PPCA models is determined when the log-likelihood reaches the maximum. Then, with regard to each local PPCA model, two monitoring statistics, i.e., Hotelling's $T$-squared ($T^2$) and squared prediction error (SPE), are calculated by utilizing the probability density of the score vector. To this end, a procedure for on-line data-driven fault diagnosis is developed for nonlinear process.

According to the procedure aforementioned, it is evidently observed that standard MPPCA has several local models and each local PPCA model has two monitoring charts.  Thus, there are excessive monitoring charts to be observed for MPPCA, which is difficult for staff to acquire the accurate information timely and quickly.

This paper is mainly dedicated to this issue, where just one or two global monitoring statistics are designed to monitor the nonlinear process. To realize this aim, $T^2$ and SPE monitoring statistics are integrated to achieve reliable monitoring performance for each local PPCA model. Then, the weighted mean of the combined monitoring statistics aforementioned in PPCA models is developed to detect underlying abnormalities.  Besides, kernel density estimation (KDE) scheme is applied to calculate thresholds in order to reduce computational overload and enhance generality. Detailed introduction about the proposed fault diagnosis approach is described below.


\subsection{Optimal number of the local models}

It can be evidently discovered that key model parameters are largely affected by the number of local models $K$. Besides, the computational cost increases with  $K$. Therefore, it is essential to seek for a reliable criteria to determine an optimal value of $K$.

In this paper, the optimal number of local models, ${K^ \circ }$, can be determined by the following criteria~\cite{yang}:
\begin{equation}\label{mn}
{K^ \circ } = \arg \mathop {\min }\limits_i H\left( i \right)
 \end{equation}
\begin{equation}\label{mm}
H\left( i \right) \equiv  - \frac{1}{N}\sum\limits_{n = 1}^N {\sum\limits_{i = 1}^K {p\left( {i|{\boldsymbol t_n},{\boldsymbol \theta} } \right)\ln \left( {p\left( {{\boldsymbol t_n}|i} \right)} \right) - \sum\limits_{i = 1}^K {{\pi _i}\ln {\pi _i}} } }
\end{equation}
where ${\boldsymbol \theta}$ is the set of  all model parameter vectors, containing ${\boldsymbol \theta}= \left\{  {\boldsymbol W_i},{\boldsymbol \mu_i},\sigma _i^2, {\pi_i} \right \}_{i = 1, ...,K}$.

Given a value of $K$, $H(i)$ is calculated after the implementation of two-stage EM schedule as the model parameter ${\boldsymbol \theta}$ is contained in $H(i)$. Thus, the model structure and parameters are embedded in each other. The value of $K$ can be determined alternatively through discovering the smallest integer satisfying $\left| {H\left( i \right) - H\left( {i + 1} \right)} \right| > \delta $, where $\delta $ is a predefined threshold.

Note that the maximum number of local models ${K_{\max }}$ varies relying on data pattern. However, ${K_{\max}}$ $=$ $5\sim10$ is typical.

\subsection{Monitoring scheme of I-MPPCA}
 I-MPPCA is proposed to solve nonlinear fault diagnosis problem, which is exactly based on a mixture of PPCA-based fault detection models.

For each $i$th local model, $T_i^2$ and $SPE_i$ are monitoring statistics for principal component subspace and residual component subspace respectively. As two subspaces are mutually orthogonal, $T^2$ statistic can not detect faults that occur in the residual component subspace and vice versa~\cite{data2014sx}. The above two monitoring statistics utilize the identical measurement unit, i.e. Mahalanobis norm, and they can be integrated into one chart. Therefore, $T_{c,i}$ is proposed based on the integration of $T_i^2$ and $SPE_i$ for each local improved PPCA model. Three monitoring statistics are computed as follows
\begin{equation}\label{t2}
{T_i^2} = {\left\| {{\boldsymbol M_i}{\boldsymbol W_i^{\rm T}}{\boldsymbol t_n}} \right\|^2}
\end{equation}
\begin{equation}\label{spe}
  SPE_i = {\left\| {{\sigma_i ^{ - 1}}(\boldsymbol I - \boldsymbol W_i{{\boldsymbol M_i}}{\boldsymbol W_i^{\rm T}}){\boldsymbol t_n}} \right\|^2}
\end{equation}
\begin{equation}\label{tc}
T_{c,i}^2 ={\boldsymbol t_n^{\rm T}}{{{({\sigma_i ^2}{\boldsymbol I} + {\boldsymbol W_i}{\boldsymbol W_i^{\rm T}})}^{ - 1}}}{\boldsymbol t_n}
\end{equation}

In the proposed approach, $R_{ni}$ is regarded as weight, which measures the degree of $n$th sample belonging to $i$th local model. Thus, the associated global monitoring statistics can be formulated as
\begin{equation}\label{t22}
 {T^2} = {{\sum\limits_{i = 1}^K {{R_{ni}}T_i^2} } \mathord{\left/
 {\vphantom {{\sum\limits_{i = 1}^K {{R_{ni}}T_i^2} } {\sum\limits_{i = 1}^K {{R_{ni}}} }}} \right.
 \kern-\nulldelimiterspace} {\sum\limits_{i = 1}^K {{R_{ni}}} }}
\end{equation}
\begin{equation}\label{spe2}
 SPE = {{\sum\limits_{i = 1}^K {{R_{ni}}SP{E_i}} } \mathord{\left/
 {\vphantom {{\sum\limits_{i = 1}^K {{R_{ni}}SP{E_i}} } {\sum\limits_{i = 1}^K {{R_{ni}}} }}} \right.
 \kern-\nulldelimiterspace} {\sum\limits_{i = 1}^K {{R_{ni}}} }}
\end{equation}
\begin{equation}\label{tc2}
  T_c^2 = {{\sum\limits_{i = 1}^K {{R_{ni}}T_{c,i}^2} } \mathord{\left/
 {\vphantom {{\sum\limits_{i = 1}^K {{R_{ni}}T_{c,i}^2} } {\sum\limits_{i = 1}^K {{R_{ni}}} }}} \right.
 \kern-\nulldelimiterspace} {\sum\limits_{i = 1}^K {{R_{ni}}} }}
\end{equation}

In consideration of the constraint that $\sum\nolimits_i {{R_{ni}}}  = 1$, the global statistics aforementioned can be much simplified without the necessity of calculating denominator.

\begin{table*}[htbp]
\begin{center}
\caption{A brief comparative study of four approaches}\label{Table0}
\renewcommand{\arraystretch}{1.1}
 \begin{tabular}{l l l l l l l}
\hline
\hline
 {\textbf{Approach}}    &\textbf{Computational cost}    & \textbf{Parameter}       &\textbf{Robustness}   &\textbf{Outliers}   & \textbf{Big data}   & \textbf{fault information}\\
\hline
K-means                 & Low: $O(N)$                   & no                       &  sensitive           & sensitive   & appropriate         & only detecting whether faults occur \\
FPCM                    & Low: $O(N)$                   & no                       &  sensitive           & insensitive  & appropriate         & only detecting whether faults occur\\
KPCA                    & High:  $O(N^2)$               & No. of PCs               &  sensitive           & insensitive  & inappropriate       &  fault detection  in two subspaces\\
I-MPPCA                 & Medium: $O(KN)$            & No. of PCs, $K$          &  insensitive         & insensitive  & appropriate         &  fault detection in two subspaces\\
\hline
\hline
\end{tabular}
\end{center}
\end{table*}

\subsection{Summary of the proposed approach}
For the fault diagnosis purpose, the thresholds under normal condition are recognized as reference to detect underlying faults. With regard to traditional PCA-based approach, thresholds are calculated under the assumption of Gaussian distributed data.
However, process variables can hardly satisfy this requirement in practical applications. In order to reduce computational cost and enhance generality, KDE technique is employed to calculate the associated thresholds~\cite{kcca}, which can be applied to both Gaussian distributed data and non-Gaussian distributed data. The basic theory is described as follows.
\begin{equation}\label{p2}
 \mathop p\limits^ \wedge  (z) = \frac{1}{{Nh}}\sum\limits_{n = 1}^N {\psi (\frac{{z - {z_n}}}{h})}
\end{equation}
where ${z_n}(n = 1, \ldots ,N)$ are the values of monitoring statistics, $h$ is the bandwidth of kernel function $\psi \left(  \cdot  \right)$. The selection of $h$ is significant because the consequences of $p(z)$ estimation would be rough if $h$ is small, whereas the density curve would be smooth. In this paper, the optimal bandwidth $h_{opt}$ is determined by minimizing the approximation of the mean integrated square error, as depicted in (\ref{hopt}), where $s$ is the standard deviation~\cite{kcca}.
\begin{equation}\label{hopt}
 {h_{opt}} = 1.06s {N^{ - 1/5}}
\end{equation}

Given a confidence level $\alpha$, the associated threshold $J_{th}$ of the monitoring statistic $J$ can be calculated by
 \begin{equation}\label{jt2}
   \int_{ - \infty }^{{J_{th}}} {p({J})d} {J} = \alpha
 \end{equation}

$J$ can be replaced by $T^2$, SPE and $T_c^2$ to acquire the corresponding threshold, namely, $J_{th,T^2}$, $J_{th,SPE}$, $J_{th,T_c^2}$. Thus, the fault detection logic follows
\begin{equation}\label{fa1}
Fault\;alarm = \left\{ {\begin{array}{*{20}{c}}
{0,}&{{T^2} \le {J_{th,{T^2}}}\,\rm{and}\,SPE \le {J_{th,SPE}}\,}\\
{1,}&{\rm{others}\quad \quad \quad \quad \quad \quad \quad \quad }
\end{array}} \right.
\end{equation}
or
\begin{equation}\label{fa2}
  Fault\;alarm = \left\{ {\begin{array}{*{20}{c}}
{0,}&{{T_c^2} \le {J_{th,T_c^2}}}\\
{1,}&{\rm{others}\quad }
\end{array}} \right.
\end{equation}

Eventually, the overall procedure of I-MPPCA approach for nonlinear data-driven process monitoring can be summarized in Table \ref{table1}.

Generally, the novel monitoring statistic $T^2_c$ by (\ref{tc}) as well as (\ref{tc2}) and the corresponding threshold are adopted to improve fault detectability, which is considerably simple and effective. As regard to the proposed approach, missing alarm rates (MARs) and false alarm rates (FARs) are mainly considered therein to evaluate the performance. It is expected that two indexes are better to approach zero.
\begin{equation}\label{mar}
\rm{MAR} = \frac{{number\;of\;samples\,\left( {\emph{J} \le  {\emph{J}_{\emph{th}}}|\emph{f}\ne 0} \right)}}{{total\;samples\,\left( {\emph{f} \ne 0} \right)}}\times 100\%
\end{equation}
\begin{equation}\label{far}
\rm{FAR} = \frac{{number\;of\;samples\,\left( {\emph{J} > {\emph{J}_{\emph{th}}}|\emph{f} = 0} \right)}}{{total\;samples\,\left( {\emph{f} = 0} \right)}}\times 100\%
\end{equation}

\subsection{Comparison with other approaches}

\begin{figure*}[htbp]
\centering
\includegraphics[width=0.98\textwidth]{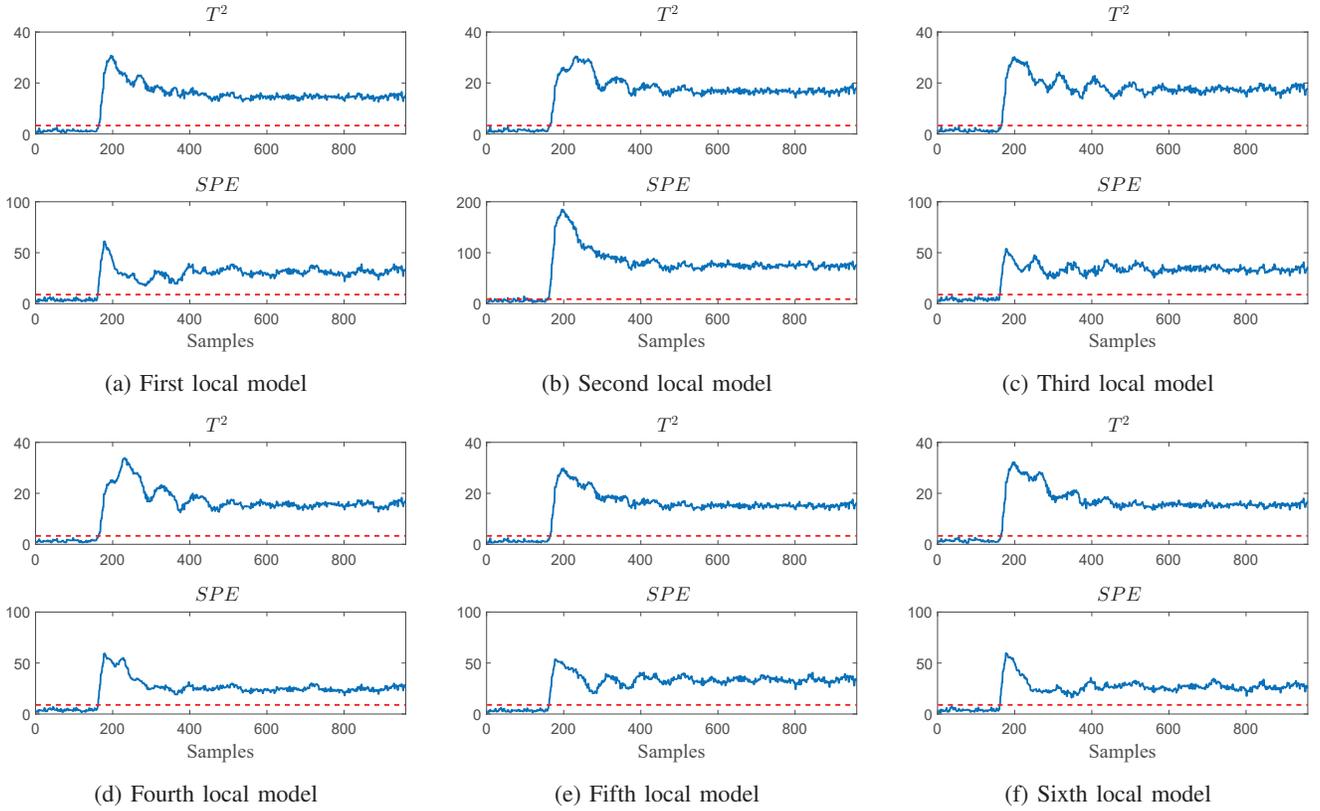}
\caption{Monitoring charts of each PPCA model using traditional MPPCA}
\label{PPCA1}
\end{figure*}

Due to the absence of data labels in most cases, several unsupervised techniques are discussed to illustrate the superior performance of I-MPPCA in this subsection. Since clustering and feature extraction techniques are typical unsupervised schemes, K-means~\cite{likas2003global,Kmeans}, fuzzy-possibilistic c-means (FPCM)~\cite{pal1997mixed} and KPCA are expected to compare with the proposed approach. The core of process monitoring is a binary classification issue. K-means and FPCM are popular classification schemes and thus can be used for fault diagnosis purpose. The number of clusters is set to be $2$.

Several performance indicators are discussed among these approaches, including  computational cost,  key parameters, robustness, etc. Notice that the computational complexity of on-line monitoring phase is critical and is valuable because this phase is implemented to monitor real-time operations. Besides, practical industry has high requirement of real-time performance.  Therefore, computational cost mainly refers to on-line monitoring procedure in this paper.

K-means is a partition scheme through seeking for certain clustering centers iteratively. It is considerably simple and easy to implement for massive data. K-means is sensitive to outliers since cluster centers are seriously influenced by outliers. Nevertheless, this approach has a high sensitivity of initial cluster centers. Once initial centers are chosen improperly, it is unlikely to acquire effective classification results. The complexity for general clustering problems is $O(kN)$, and $k$ is the number of clusters, with $k=2$ for fault diagnosis. Thus, the complexity for process monitoring is $O(N)$.

FPCM has more preferable reliability than K-means. The major spirit of FPCM is to acquire the membership vector via minimizing the objective function.  The classification consequences are also sensitive to initial cluster centers.
FPCM can solve the noise sensitivity of defect~\cite{pal1997mixed}. However, it is unable to detect which subspace the fault occurs.

The technical core of KPCA is to map low-dimensional data into high-dimensional linear space and PCA is performed in the high-dimensional feature space. However, KPCA is sensitive to parameter tuning, especially the kernel bandwidth. The computational complexity is $O(N^2)$ owing to the calculation of Gaussian kernels, which makes it inappropriate for real-time process monitoring. In addition, as PCA is insensitive to outliers, this property is inherent in KPCA and I-MPPCA.

I-MPPCA approach partitions the data into several models via the technique of a mixture of probabilistic principal component analysers. It can deal with missing data while the other three approaches are lack of this property~\cite{wagstaff2004clustering,sanguinetti2006missing}.
Two key parameters, the number of local models $K$ and the number of principal components (PCs), are simply determined with less computational cost. The computational complexity of the proposed approach is $O(KN)$. In practical applications, it is obvious that $K \ll N$, which indicates that I-MPPCA is less complicated than KPCA and suitable to process large data.
Besides, it is insensitive to parameter tuning since $K$ has limited impact on monitoring consequences when $K$ varies in a certain range. Moreover, I-MMPCA can estimate which subspace faults occur by selecting appropriate monitoring statistic $T^2$ or SPE.

Major characteristics are concluded in Table \ref{Table0}, which implies that I-MPPCA is superior to the others to some extent.

\section{Case study on TE process}\label{TE_CASE}
The proposed I-MPPCA is exemplified through the TE process in this section.  Local improved PPCA models are embedded in traditional MPPCA framework and global monitoring statistic is calculated. This case study is utilized to evaluate the rationality and superiority of I-MMPCA approach in comparison with traditional MPPCA approach.  The results are obtained through simulations on MATLAB.


TE process model is a realistic chemical plant simulator that serves as a preferred benchmark for monitoring study~\cite{TE1,TE2}. Since prior knowledge about the mathematical model of TE process is unavailable, the monitoring approach can be designed only based on sensing measurements. $20$ process faults were initially defined and are adopted in this study, namely, IDV(1)-IDV(20). More detailed introduction was described in~\cite{TE}.

\begin{figure*}[htbp]
\centering
\includegraphics[width=0.98\textwidth]{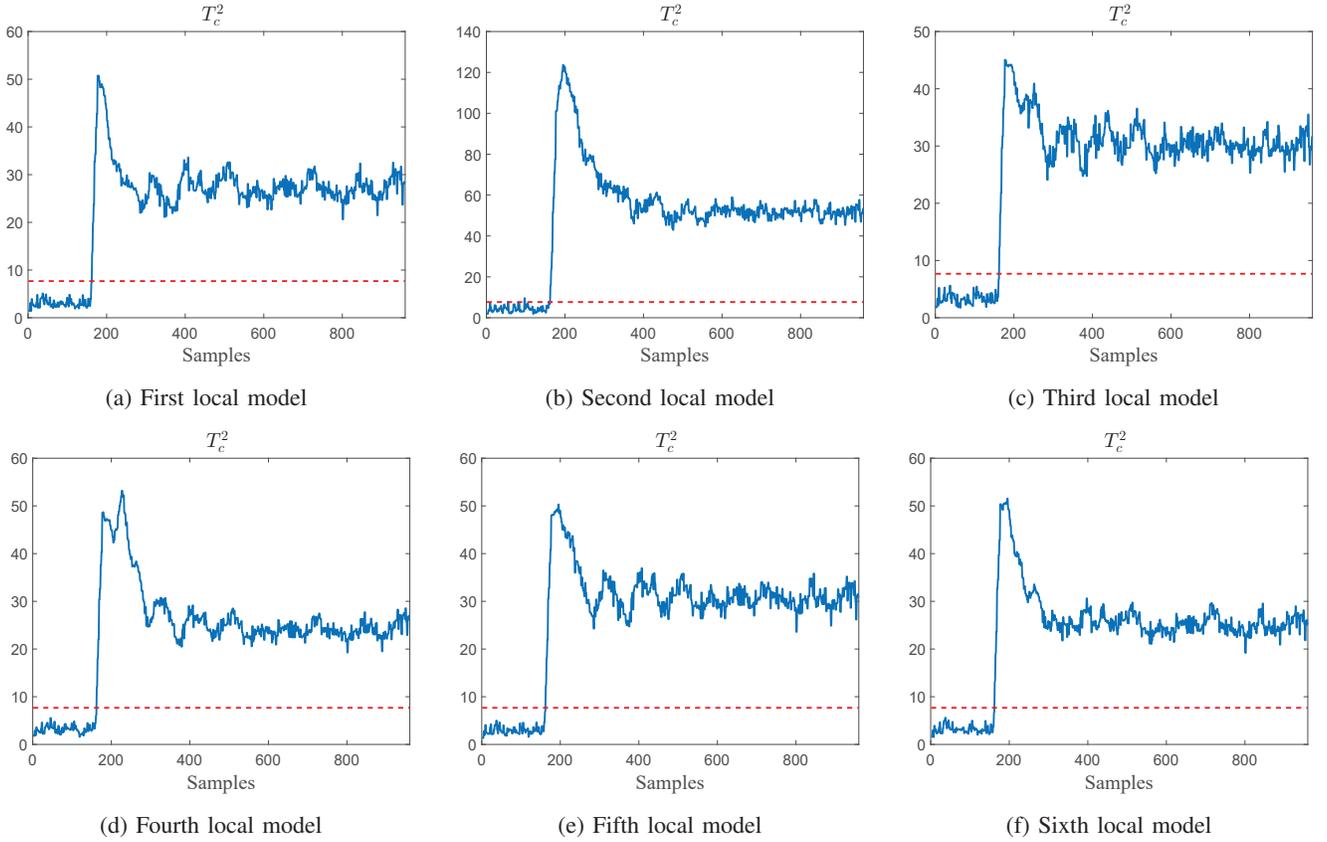}
\caption{Monitoring charts of each local improved PPCA model under traditional MPPCA framework}
\label{CPPCA1}
\end{figure*}

\begin{figure*}[htbp]
\centering
\includegraphics[width=0.98\textwidth]{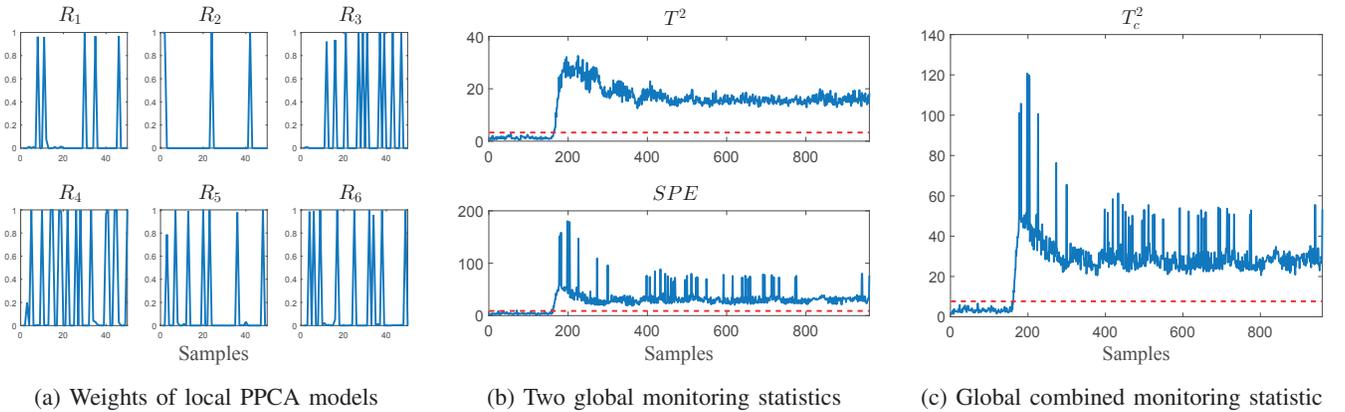}
\caption{Monitoring charts using I-MPPCA approach}
\label{IMPPCA1}
\end{figure*}

\begin{table*}[htbp]
\caption{MARs (\% ) based on TE data given in \cite{TE}}\label{Table2}
\centering
\renewcommand{\arraystretch}{1.2}
\sloppy
\begin{tabular}{c c c c c c c c c c c c}
\hline
\hline
Fault         &IDV(1)  &IDV(2) &IDV(3)  &IDV(4) &IDV(5)  &IDV(6)  &IDV(7)  &IDV(8) &IDV(9) &IDV(10)\\
\hline
MPPCA ($T^2$ or SPE)      & 0.75   &1.63   &83.88   & 6.88  &7.75   &0       &0       &1.25   &84.37  &39.12  \\
I-MPPCA ($T^2_c$)        &0       &0.63   &83.63   &0.5    &6.13   &0       &0       &0.88   &83.75  &36.63 \\
\hline
Fault        & IDV(11) &IDV(12) &IDV(13) &IDV(14) &IDV(15) &IDV(16) &IDV(17)  &IDV(18) &IDV(19) &IDV(20) \\
\hline
MPPCA ($T^2$ or SPE)    &53.50    &3.50    &3.62    &21.88   &84.25   &16.88   &36.88    &9.00   &84.00   &18.50   \\
I-MPPCA ($T^2_c$)      &56.87    &1.75    &3.50    &12.13   &63.50   &16.13   &36.00    &8.38   &87.75   &16.75  \\
\hline
\hline
\end{tabular}
\end{table*}

In this simulation, $22$ control variables and $11$ manipulated variables are chosen as the samples. $960$ normal samples are utilized to acquire off-line learning model. $960$ testing samples, including the first $160$ normal samples and $800$ subsequent faulty samples, are adopted to evaluate the performance.  The confidence level is set to be $0.99$. The number of PCs is selected as $6$ based on accumulation contribution rate. According to the criterion described by (\ref{mn}) and (\ref{mm}), the number of local PPCA models is $6$.

Then, fault IDV(1) is adopted to illustrate the rationality of I-MPPCA approach.
Detailed monitoring consequences of traditional MPPCA approach are shown in Fig.\ref{PPCA1}, which are calculated by (\ref{t2}) and (\ref{spe}). With regard to each local improved PPCA model, another monitoring statistic based on the integration of $T^2$ and SPE is described by (\ref{tc}), as illustrated in Fig.~\ref{CPPCA1}. According to Fig.~\ref{PPCA1} and  Fig.~\ref{CPPCA1}, fault can be detected timely and accurately by $12$ or $6$ monitoring graphs, respectively.

\begin{figure}[!bp]
\centering
\includegraphics[width=0.48\textwidth,angle=0]{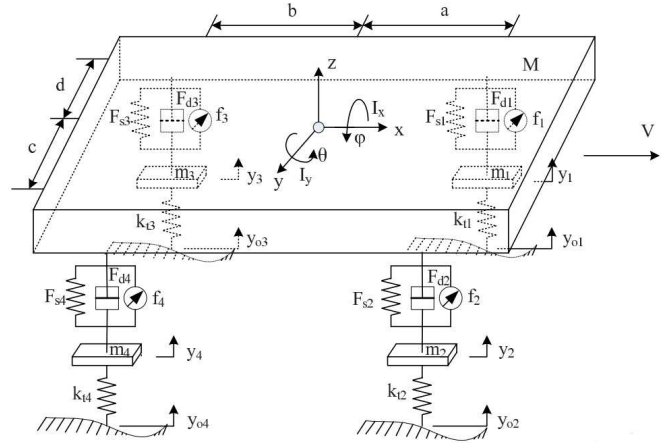}
\caption{Flow diagram of autosuspension model}
\label{benchmark1}
\end{figure}

\begin{figure*}[!htp]
\centering
\includegraphics[width=1\textwidth]{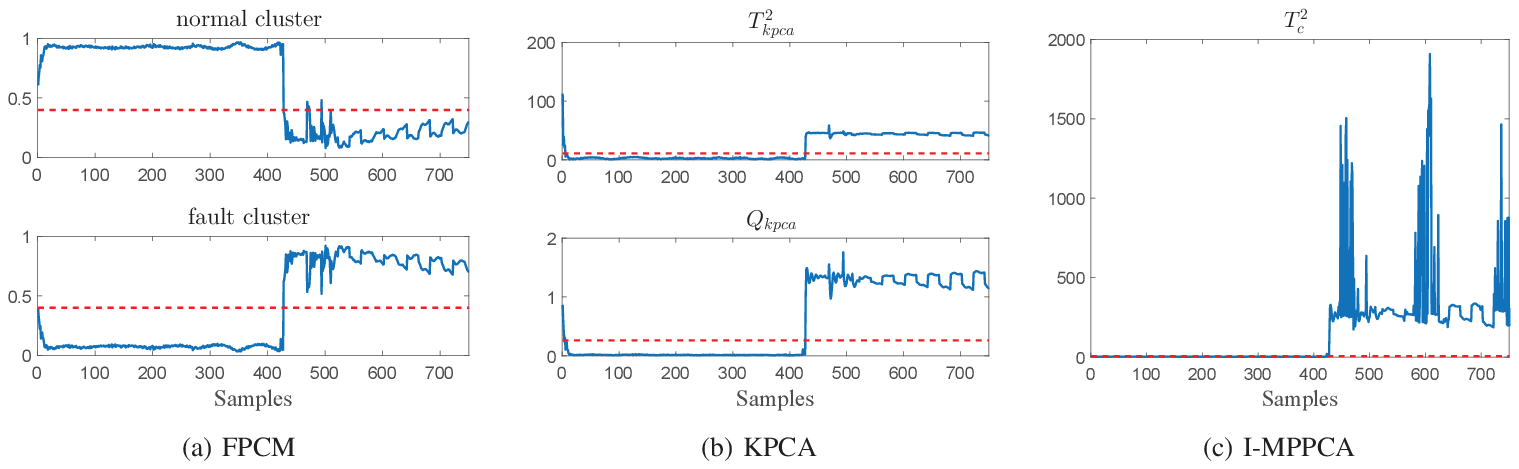}
\caption{Monitoring charts of Fault 1}
\label{F1}
\end{figure*}

\begin{figure*}[!htp]
\centering
\includegraphics[width=1\textwidth]{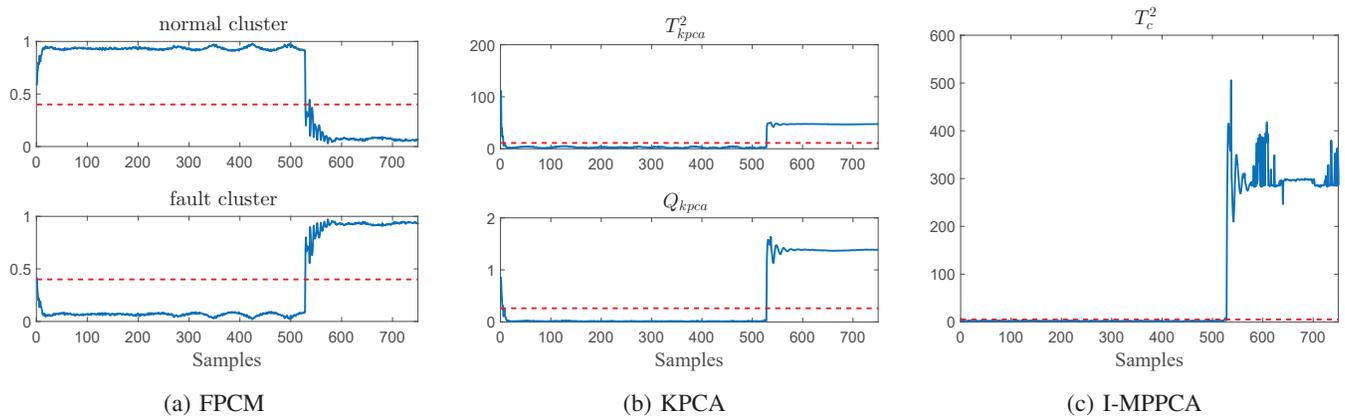}
\caption{Monitoring charts of Fault 2}
\label{F2}
\end{figure*}

The monitoring consequences of I-MPPCA are represented in Fig. \ref{IMPPCA1}.  Posterior  probability $R$ is regarded as weight, which measures membership degree a data point belonging to a certain local model of being chosen. In order to observe clearly, weights of the first $50$ samples are revealed in Fig. 3a. The weights are almost $1$ or $0$, which indicates that the samples belong to the certain local PPCA model completely or not, and proves the rationality of the proposed global monitoring statistics. And the monitoring charts of statistics (\ref{t22})-(\ref{spe2}) are demonstrated in Fig. 3b and the statistic (\ref{tc2}) is illustrated in Fig. 3c. The global monitoring statistics may change sharply at several points because the weights of the second local model are nearly $1$ and the corresponding values of SPE as well as $T_c^2$ are relatively larger than those of other local models.
According to the comparison among Fig.~\ref{PPCA1}, Fig.~\ref{CPPCA1}, Fig. 3b and Fig. 3c, the accuracy rates of $4$ sorts of calculation approaches are almost close. However, the number of monitoring charts would be reduced from $12$ to $1$ gradually, which is exactly the major advantage of I-MPPCA approach.

In addition, faults IDV(1)-IDV(20) are employed to demonstrate the superior performance of the global combined statistic in respect of false alarm rates and missing alarm rates.
The detailed MARs are listed in Table~\ref{Table2}. It can be evidently acquired that the MARs of $T^2_c$ are almost smaller than those of $T^2$ or SPE.
Besides, the FAR of $T^2$ or SPE is $7.5\%$ while the FAR of $T^2_c$ is $2.5\%$.  Moreover, the computational complexity is almost similar. Therefore, I-MPPCA approach provides significant improvement compared with traditional MPPCA technique in terms of accuracy and convenience.

\section{Case study on Autosuspension benchmrk}\label{autosuspension_case}
In this section, autosuspension benchmark is employed to demonstrate the superiority through comparison with K-means, FPCM and KPCA. The basic description of autosuspension model is present briefly in Subsection~\ref{autosuspension}. Then, complete data and missing data cases are considered to illustrate the optimal performance of I-MPPCA even when some data values are missing, respectively.

\subsection{Brief description of Autosuspension system}\label{autosuspension}

\begin{table}[!bp]
\begin{center}
\caption{The parameters of the autosuspension model}\label{table5}
\renewcommand{\arraystretch}{1.1}
\begin{tabular}{ll}
\hline
\hline
Notation              & Description\\
\hline
$M$                   & Vehicle body mass\\
${m_i}$               & Unsprung mass\\
$\theta$              &Pitch motion\\
$\varphi$             & Roll motion\\
${I_x}$               & Roll motion rotary inertias \\
${I_y}$               & Pitch motion rotary inertias \\
${y_i}$               & Unsprung mass displacement\\
$\Delta {y_i}$        & Suspension deflection \\
$\Delta {\dot{y_i}}$  & Deflection velocity\\
${y_{0i}}$            & Road input\\
${k_i}$               & Linear stiffness parameter \\
${k_{ni}}$            & Nonlinear stiffness parameter\\
${b_{ei}}$            & Extension movement damping parameter\\
${b_{ci}}$            & Compression movement damping parameter\\
${k_{ti}}$            & Stiffness of the tire\\
${F_{si}}$            & Force produced by the spring \\
 ${F_{di}}$           & Force produced by the damper\\
 ${f_i}$              & Force produced by the related actuator\\
$a,b,c,d$             & Distances \\
\hline
\hline
\end{tabular}
\end{center}
\end{table}

The basic flow diagram of this model is presented in Fig.~\ref{benchmark1} and a comprehensive introduction was described in~\cite{vehicle}. Besides, the parameters are listed in Table~\ref{table5} and specific values can be found in~\cite{vehicle}.

To our best knowledge, the front wheels (suspension 1 and suspension 2) have the identical configuration and the rear wheels (suspension 3 and suspension 4) share a different one.  In other words, researchers just need to study suspension $1$ and suspension $3$.
In practice, most common faults originate from the aging of suspension components, for instance, the parameter reductions from spring and damper. It is difficult to establish specific mathematical model owing to lack of sufficient process knowledge. Therefore, it is essential to implement data-driven techniques for autosuspension monitoring.

Several sorts of sensors are available in industrial applications, e.g., laser sensor, accelerometers, grometer and linear variable displacement transducer. For process monitoring task, only accelerometers are useful and sensing measurements from four accelerometers are adopted in this study. That is, the dimension of data is $4$.

\subsection{Simulation with complete data}\label{complete}

In this section, suspension coefficient reduction is taken as an example in this paper.
These techniques can also be applied to damper coefficient reduction case. $750$ normal samples are generated to train I-MPPCA model or KPCA model. The threshold of FPCM should be calculated by normal samples. K-means need not to train model in advance. Then, $750$ testing samples are generated as follows:

 1) Fault 1, the spring coefficient of suspension 1 is reduced by $30\%$ from the $429$th sample;

 2) Fault 2, the spring coefficient of suspension 3 is reduced by $30\%$ from the $529$th sample.

Since K-means just provides two classification labels, only the accuracy rates (MARs and FAR) are given in this paper. Therefore, the monitoring charts of the other three approaches are illustrated in Fig. \ref{F1} and Fig. \ref{F2}. It can be evidently observed that FPCM, KPCA and I-MPPCA can detect faults timely.

\begin{figure*}[htp]
\centering
\includegraphics[width=1\textwidth]{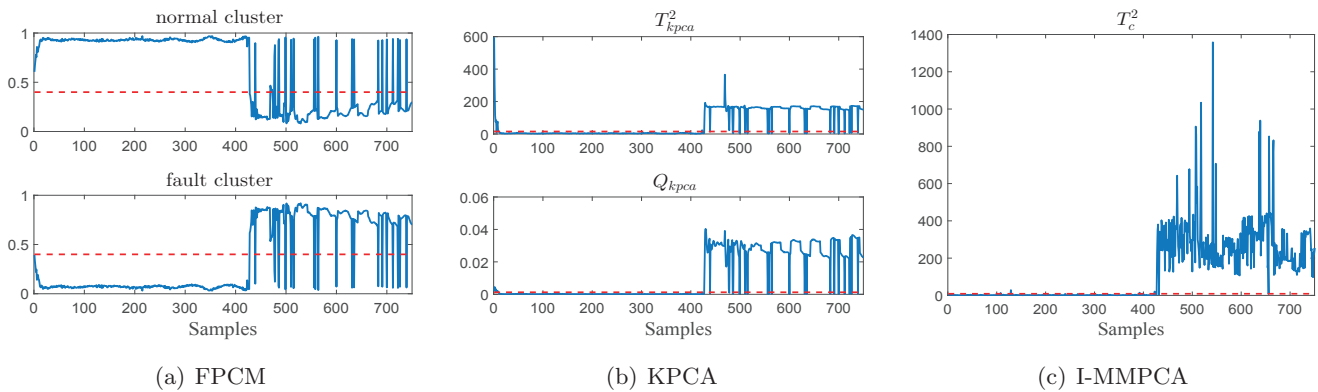}
\caption{Monitoring charts of Fault 3}
\label{fault1-5}
\end{figure*}

\begin{figure*}[htp]
\centering
\includegraphics[width=1\textwidth]{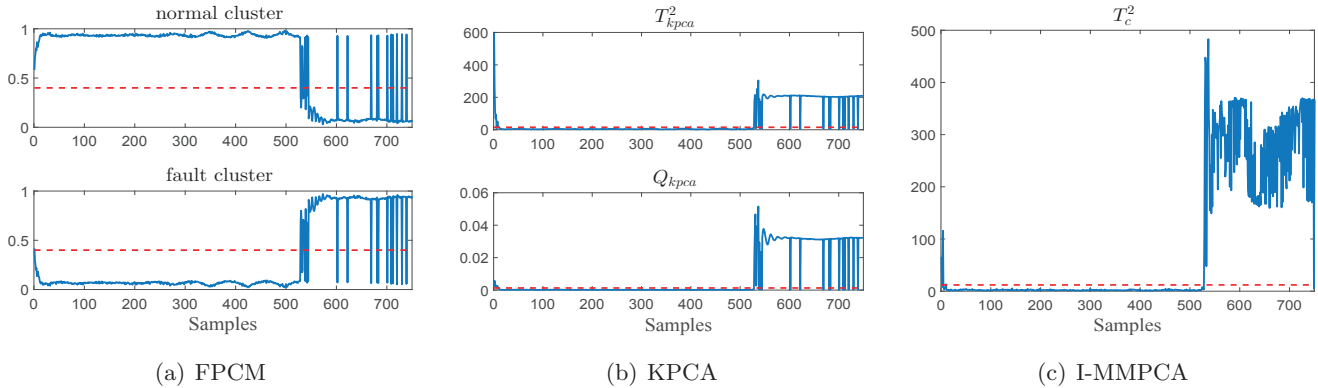}
\caption{Monitoring charts of Fault 4}
\label{fault2-5}
\end{figure*}

With regard to FPCM, there would be some misleading places where two membership values are both larger than the predefined threshold. Besides, FPCM can not detect which subspace faults happen.
KPCA has two monitoring statistics for residual component subspace and principal component subspace. And the consequences reveal that faults appear in both subspaces. For I-MPPCA approach, it enables providing this information when global $T^2$ and SPE statistics are selected.

Besides, the MARs and FARs are listed in Table \ref{tb1} and Table \ref{tb2}, respectively.
For this simulation case, KPCA and I-MPPCA can discover faults completely accurately while K-means and FPCM have several missing alarm points. Moreover, the FAR of I-MPPCA approaches to zero, lower than those of K-means and KPCA. Generally speaking, I-MPPCA delivers optimal monitoring performance among four techniques through the trade-off of FARs and MARs.

In conclusion, I-MPPCA has relatively higher detection accuracy rates based on the analysis above.

\subsection{Simulation with incomplete data}

In order to compare conveniently, autosuspension data generated in Subsection~\ref{complete} with different artificial missing schemes are employed to demonstrate that I-MPPCA can deliver optimal performance when some data values are missing.

Note that missing data values are generated randomly to simulate the practical systems. Besides, with regard to the proposed approach, the tolerant maximal missing rate for modeling data is $30\%$ in this paper and the training accuracy rates basically remain the same with the increasing missing rates of modeling data before the tolerant maximal missing rate. Therefore, 750 training samples with 15\% missing data are taken as an example and utilized to establish model in this study.  Then, testing samples with missing data values are generated as follows:

1) Fault 3, the spring coefficient of suspension 1 is reduced by $30\%$ from the $429$th sample with 5\% missing data;

2) Fault 4, the spring coefficient of suspension 3 is reduced by $30\%$ from the $529$th sample with 5\% missing data;

3) Fault 5, the spring coefficient of suspension 1 is reduced by $30\%$ from the $429$th sample with 10\% missing data;

4) Fault 6, the spring coefficient of suspension 3 is reduced by $30\%$ from the $529$th sample with 10\% missing data.

MARs and FARs are concluded in Table~\ref{tb1} and Table~\ref{tb2}. As to I-MPPCA approach, it can be observed clearly that missing data values have the least influence on monitoring performance and the FAR as well as MARs are basically the lowest. Especially, according to the detection accuracy  rates of $6$ faults, the accuracy rates will be reduced rapidly with the increasing missing rate.

\begin{figure*}[htbp]
\centering
\includegraphics[width=1\textwidth]{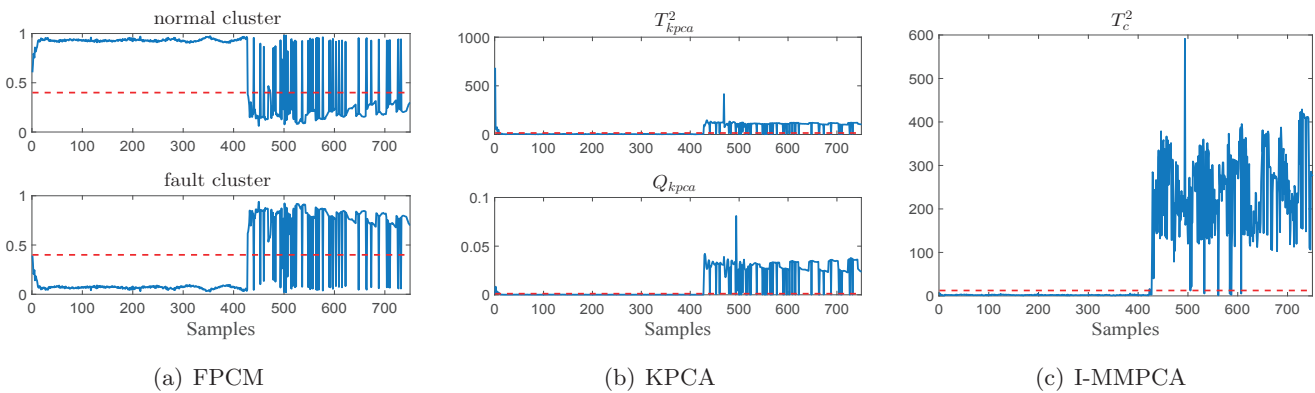}
\caption{Monitoring charts of Fault 5}
\label{fault1-10}
\end{figure*}

\begin{figure*}[!htbp]
\centering
\includegraphics[width=1\textwidth]{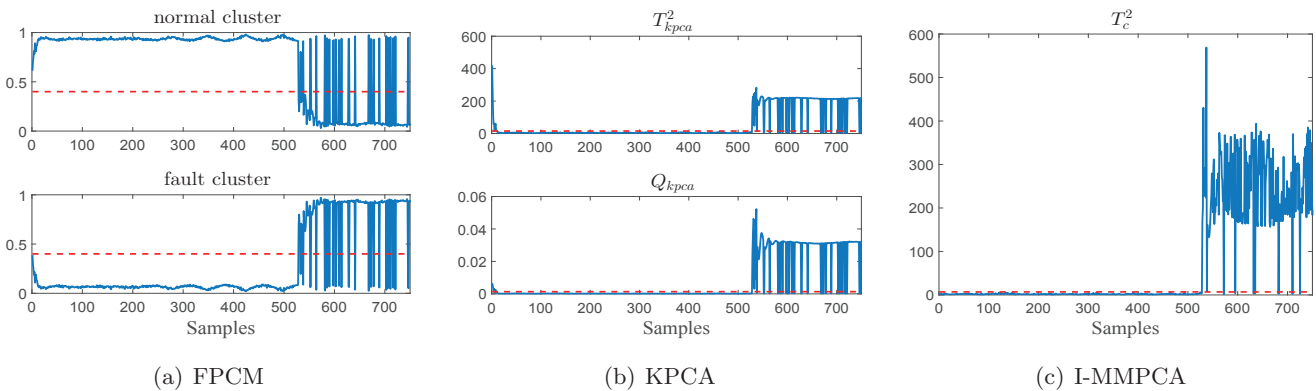}
\caption{Monitoring charts of Fault 6}
\label{fault2-10}
\end{figure*}

Detailed monitoring charts of four faults are shown in Figs.\ref{fault1-5}-\ref{fault2-10}.
According to the comparison of Fault $1$, Fault $3$ and Fault $5$, Fault $2$, Fault $4$ and Fault $6$, it can be seen that missing rate has important effect on K-means, FPCM and KPCA. Besides, the missing alarm points are exactly the positions of missing samples. And MARs are seriously influenced by which variable occurs missing data. However, the FAR and MARs of I-MPPCA approach are slightly affected by these missing data values.

In conclusion, with regard to missing data, I-MPPCA delivers optimal performance by comparison with three approaches. Furthermore, the proposed approach owns extra virtues, for instance, low computational complexity, insensitivity to parameter tuning, being able to monitor two subspaces, etc.  Therefore, I-MPPCA is superior to the others and prior for nonlinear process monitoring both in practical applications and in academic study.

\begin{table}[!tbp]
\caption{MARs (\%) based on autosuspension data}
\centering
\renewcommand{\arraystretch}{1.1}
\begin{tabular}{c c  c c c}
\hline
\hline
Approach   &K-means    &FPCM    &KPCA       & I-MPPCA\\
\hline
Fault 1    &2.82       &1.41    &0          &\textbf{0}\\
Fault 2    &2.85       &1.90    &0          &\textbf{0}\\
Fault 3    &7.14       &6.52    &6.21       &\textbf{0.62}\\
Fault 4    &8.11       &7.66    &6.76       &\textbf{0.45}\\
Fault 5    &13.66      &13.04   &12.11      &\textbf{1.86}\\
Fault 6    &15.77      &13.96   &13.06      &\textbf{3.15}\\
\hline
\hline
\end{tabular}
\label{tb1}
\end{table}

\begin{table}[!tbp]
\caption{FARs (\%) based on autosuspension data}
\centering
\renewcommand{\arraystretch}{1.1}
\begin{tabular}{c c  c c c}
\hline
\hline
  Approach        &K-means    &FPCM    &KPCA    & I-MPPCA\\
\hline
  Complete data   &2.82       &0.18       &1.41    & \textbf{0}\\
  Missing data    &5.64        &1.04   &2.57    & \textbf{0.47}\\
\hline
\hline
\end{tabular}
\label{tb2}
\end{table}

\section{Conclusion}\label{conclusion}

This paper has proposed  a novel computing method of monitoring statistics under the framework of traditional MPPCA for nonlinear data-driven process monitoring. Appropriate partitioning of sensing measurements and the parameters of local PPCA models are automatically acquired via the technique of a mixture of probabilistic principal component analysers. Besides, a two-stage EM schedule is employed to improve the convergence speed and reduce computational cost. A novel composite  monitoring statistic has been introduced and calculated in each PPCA model aforementioned. It is shown that the posterior probability can be regarded as a weight to  data point belonging to a certain PPCA model of being chosen. Therefore, in order to provide an optimal fault detection performance and observational convenience, the global monitoring statistics are acquired based on the weighted mean of all local monitoring statistics. Moreover, several typical unsupervised schemes including feature extraction algorithms and clustering approaches have been discussed to highlight the virtues of the I-MPPCA including low computational cost,  parameter robustness, the capability of dealing with incomplete data, etc. Finally,   simulation  studies  in comparison with  several known approaches have been carried out based on TE data and an autosuspension model which have demonstrated the superior performance of the proposed approach.

\appendices

\section{Solution procedure of MPPCA}\label{mppcasolution}
According to (\ref{pt}), the log-likelihood of observation data for a mixture model can be depicted as:
\begin{equation}\label{eq11}
\begin{array}{l}
L = \sum\limits_{n = 1}^N {\ln \left\{ {p\left( {{\boldsymbol t_n}} \right)} \right\}} \\
\;\;{\kern 1pt}  = \sum\limits_{n = 1}^N {\ln \left\{ {\sum\limits_{i = 1}^K {{\pi _i}} p\left( {{\boldsymbol t_n}|i} \right)} \right\}}
\end{array}
\end{equation}

The maximum-likelihood is utilized to determine key parameters of the model, where the proper segmentation of the data occurs automatically when the log-likelihood reaches its maximum. An iterative EM algorithm  is developed to optimize the model parameters $\pi _i$, $\boldsymbol \mu_i$, $\boldsymbol W_i$ and $\sigma _i^2$, which was first introduced in \cite{iterativeEM}.

 Suppose that ${R_{ni}} = p\left( {i|{\boldsymbol t_n}} \right)$ is the posterior probability of the $i$th local model for generating data $\boldsymbol t_n$, it can be estimated using Bayesian rule
\begin{equation}\label{rni}
  {R_{ni}} = \frac{{p\left( {{\boldsymbol t_n}|i} \right){\pi _i}}}{{p\left( {{\boldsymbol t_n}} \right)}}
\end{equation}

 With regard to this posterior distribution, the expectation of $L_C$ can be acquired in the form of
\begin{align}\label{lc}
&\left\langle {{L_c}} \right\rangle=\sum_{n=1}^{N} \sum_{i=1}^{K} {R_{ni}} \left\{ \ln {\pi _i}-\frac{d}{2}\ln \sigma _i^2- \frac{1}{2} tr\left\{ \left\langle   {{\boldsymbol x^{(i)}_{n}}}  {{\boldsymbol x^{(i)}_{n}}}^{\rm T}   \right\rangle \right\}\right. \nonumber\\
&\;\;\;\;\;\;\;\;\;\;\; - \frac{1}{{2\sigma _i^2}}{\left\| {{\boldsymbol t_{n}} - {\boldsymbol \mu _i}} \right\|}^2+ {{\frac{1} {\sigma _i^2}}} {\left\langle {{\boldsymbol x^{(i)}_{n}}} \right\rangle }^{\rm T} {\boldsymbol W_i^{\rm T}}({\boldsymbol t_{n}} - {\boldsymbol \mu _i}) \nonumber\\
&\;\;\;\;\;\;\;\;\;\;\;\left. {- \frac{1}{{2\sigma _i^2}}} tr\left\{  {\boldsymbol W_i^{\rm T}} {\boldsymbol W_i}  \left\langle   {{\boldsymbol x^{(i)}_{n}}}  {{\boldsymbol x^{(i)}_{n}}}^{\rm T}   \right\rangle  \right\} \right\}
\end{align}

In order to obtain the optimal values of key model parameters aforementioned, a Lagrange multiplier $\lambda$ is utilized to achieve the maximum value of (\ref{lc}). Thus, the solution of maximum likelihood can be transformed into the following optimization problem
\begin{equation}\label{eq17}
\left\{ \begin{array}{l}
\max \;\;\left\langle {{L_C}} \right\rangle  + \lambda \left( {\sum\limits_{i = 1}^K {{\pi _i} - 1} } \right)\\
s.t.\quad \sum {{\pi _i} = 1}
\end{array} \right.
\end{equation}


To our best knowledge, traditional EM algorithm is considerably complicated due to iterative convergence process. In this paper, a two-stage EM schedule is adopted, where generalized EM (GEM) is utilized in M-step to improve convergence speed and reduce computational complexity \cite{GEM}. The two-stage EM algorithm for MPPCA is described in detail in Appendix \ref{solutionofmppca}.

\section{A two-stage EM for MPPCA}\label{solutionofmppca}
The log-likelihood function we expect to maximize is described as the likelihood (\ref{eq11}).

 The relevant expected complete-data log-likelihood is interpreted as
 \begin{equation}\label{eq22}
 {\widehat L_C} = \sum\limits_{n = 1}^N {\sum\limits_{i = 1}^K {{R_{ni}}\ln \left\{ {{\pi _i}p({\boldsymbol t_n},{\boldsymbol x^{(i)}_{n}})} \right\}} }
 \end{equation}
where $R_{ni}$  is calculated by (\ref{rni}).  The first stage of the two-stage EM schedule (E-step) is maximizing (\ref{eq22}) to acquire ${\widetilde {\boldsymbol \mu} _i}$ and ${\widetilde {\pi }_i}$.

The second stage (M-step) takes advantage of generalised EM (or GEM) to update $\boldsymbol W_i$ and $\sigma _i^2$. The typical feature of GEM is to increase the value of ${\widehat L_C}$ and not to maximize it during the iteration process. Regarding ${\widehat L_C}$ as the likelihood of interest, one cycle of EM is performed about $\boldsymbol W_i$ and $\sigma _i^2$.

 M-step procedure can be simplified further when (\ref{lc}) is expanded for $\left\langle {{\boldsymbol x^{(i)}_{n}}} \right\rangle $ and $\left\langle {{\boldsymbol x^{(i)}_{n}}{\boldsymbol x^{(i)}_{n}}^{\rm T} } \right\rangle$, only terms in $\widetilde {\boldsymbol \mu }_i$ appear.  Thus, the expected complete-data log-likelihood now can be obtained by inspection of (\ref{lc}) as follows
\begin{align}\label{eq16}
&\left\langle {{L_c}} \right\rangle=\sum_{n=1}^{N} \sum_{i=1}^{K} {R_{ni}} \left\{ \ln {{\widetilde \pi }_i}-\frac{d}{2}\ln \sigma _i^2- \frac{1}{2} tr\left\{ \left\langle   {{\boldsymbol x^{(i)}_{n}}}  {{\boldsymbol x^{(i)}_{n}}}^{\rm T}   \right\rangle \right\}\right. \nonumber\\
&\;\;\;\;\;\;\;\;\;\;\; - \frac{1}{{2\sigma _i^2}} {\left\| {{\boldsymbol t_{n}} - {{\widetilde {\boldsymbol \mu} _i}}} \right\|}^2+ {\frac{1} {\sigma _i^2}} {\left\langle {{{\boldsymbol x^{(i)}_{n}}}} \right\rangle}^{\rm T} {\boldsymbol W_i^{\rm T}}({\boldsymbol t_{n}} - {{\widetilde {\boldsymbol \mu} _i}}) \nonumber\\
&\;\;\;\;\;\;\;\;\;\;\;\left. {- \frac{1}{{2\sigma _i^2}}} tr\left\{  {\boldsymbol W_i^{\rm T}} {\boldsymbol W_i}  \left\langle   {{\boldsymbol x^{(i)}_{n}}}  {{\boldsymbol x^{(i)}_{n}}}^{\rm T}   \right\rangle  \right\} \right\}
\end{align}

Much simplified M-step formulas can be acquired when (\ref{eq16}) reaches the maximum with respects to $\boldsymbol W_i$ and $\sigma_i^2$ (keeping ${\widetilde {\boldsymbol \mu }_i}$ fixed)
\begin{equation}\label{eq118}
  {\widetilde{\boldsymbol W}_i} = {\boldsymbol S_i}{\boldsymbol W_i}{\left( {{\sigma _i}^2{\boldsymbol I} + \boldsymbol M_i^{ - 1}{\boldsymbol W_i^{\rm T}}{\boldsymbol S_i}{\boldsymbol W_i}} \right)^{ - 1}}
 \end{equation}
 \begin{equation}\label{eq19}
  {\widetilde {{\sigma _i}}^2} = \frac{1}{d}tr\left( {{\boldsymbol S_i} - {\boldsymbol S_i}{\boldsymbol W_i}{\boldsymbol M_i^{ - 1}}{\widetilde{\boldsymbol W}_i} ^{\rm T}} \right)
 \end{equation}
 where
 \begin{equation}\label{eq20}
   {\boldsymbol S_i} = \frac{1}{{{\widetilde{\pi}_i}N}}\sum\limits_{n = 1}^N {{R_{ni}}({\boldsymbol t_{n}} - {{{\widetilde {\boldsymbol \mu }}_i}}){{({\boldsymbol t_{n}} - {{{\widetilde {\boldsymbol \mu }}_i}})}^{\rm T}}}
 \end{equation}
 \begin{equation}\label{eq23}
 { \widetilde {\boldsymbol \mu} _i} = \frac{{\sum\limits_{n = 1}^N {{R_{ni}}{\boldsymbol t_n}} }}{{\sum\limits_{n = 1}^N {{R_{ni}}} }}
 \end{equation}
\begin{equation}\label{eq18}
{  \widetilde{\pi}_i} = \frac{1}{N}\sum\limits_{n = 1}^N {{R_{ni}}}
\end{equation}
Obviously, the symbol $\;\widetilde {}\;$ indicates new variab+les that may be updated in the M-step.
 Iteration of (\ref{eq20})-(\ref{eq18}) as well as (\ref{rni}) followed by (\ref{eq118}) and (\ref{eq19}) in turns is ensured to reach a local maximum of the likelihood (\ref{eq11}).


\end{document}